\documentstyle[sprocl]{article}
\def\Journal#1#2#3#4{{#1} {\bf #2}, #3 (#4)}

\def\NPB{{\em Nucl. Phys.} B}
\def\PLB{{\em Phys. Lett.}  B}
\def\PRL{\em Phys. Rev. Lett.}
\def\PRD{{\em Phys. Rev.} D}

\def\PTP{\em Prog. Theor. Phys.}

\def\be{\begin{equation}}
\def\ee{\end{equation}}
\def\bea{\begin{eqnarray}}
\def\eea{\end{eqnarray}}

\begin{document}
\begin{titlepage}

\vspace{0.2in}
\rightline{\vbox{\halign{&#\hfil\cr
&NTUTH-97-03\cr
&June 1997\cr}}}

\vfill

\begin{center}
{\Large \bf  PERSPECTIVE ON  FCNC: \\
FROM RARE TO WELL-DONE
\footnote
{
Talk presented at FCNC97 Workshop, Santa Monica, USA, February 19--21, 1997. 
To appear in Proceedings.
}
}
\vfill
        {\bf George Wei-Shu HOU}
\footnote{
E-mail: wshou@phys.ntu.edu.tw.} \\

        {Department of Physics, National Taiwan University,}\\
        {Taipei, Taiwan 10764, R.O.C.}\\
\end{center}
\vfill
\begin{abstract}
We briefly review the phenomenology of FCNC,
from the very rare $\mu/K$, through the medium $b$, 
to the ``pseudo-well-done" case of $b^\prime$, 
extending to the possibility of large tree level
FCNC at weak scale.
\end{abstract}
\vfill
\end{titlepage}
\title{PERSPECTIVE ON  FCNC: FROM RARE TO WELL-DONE}
\author{GEORGE W.S. HOU}
\address{Department of Physics, National Taiwan University \\
       Taipei, Taiwan 10764, R.O.C. }
\maketitle\abstracts{
We briefly review the phenomenology of FCNC,
from the very rare $\mu/K$, through the medium $b$, 
to the ``pseudo-well-done" case of $b^\prime$, 
extending to the possibility of large tree level
FCNC at weak scale.
}

\section{Overview: From Down-Up}
 
In a multi-flavored world, why is FCNC so rare?
The GIM mechanism answered this definitively
and became part of the Standard Model (SM): There is no tree level FCNC,
while there is unitary cancellation at loop level.
Thus, {\bf FCNC is loop-induced and rare!}

A second problem arises from the Higgs sector:
In a multi-flavored world, why not multi-Higgs as well?
The problem is one would again have 
flavor changing neutral Higgs couplings (FCNH) at tree level.
These are removed by imposing 
the Natural Flavor Conservation (NFC) condition,\cite{GW}
usually via discrete symmetries:
``each type of fermion charge has {\it only one} source of mass",
which is an SM-like feature.
Thus, for $N_F>1$, tree level FCNC is killed by GIM,
while for $N_F,\ N_H > 1$, tree level FCNH is killed by NFC.

The context for our study of FCNC phenomenology is therefore:

\noindent (a)  MSM (Minimal SM), with heavy top as the (GIM breaking) loop-driver.

\noindent (b) MBSM (Minimal Beyond SM), i.e. minimal variations:

\noindent \ \ \ \ \ $\bullet$ Sequential fermions 
(SM4: $b^\prime$, $t^\prime$; $L^-$, $L^0$, with $m_{L^0} > M_Z/2$), or 

\noindent \ \ \ \ \ $\bullet$ Extra Higgs doublet with NFC 
(2HDM: Model I and II ($\leftarrow$ MSSM!)).

\noindent (c) Variation on the theme: Liberation from 
GIM or NFC at high energy? 

\noindent \ \ \ \, $\hookrightarrow$ Tree level FCNC at weak scale!?

Thus, starting from low energy FCNC which are 
rather GIM/NFC suppressed, 
we explore the theme that they go 
from rare to perhaps ``well-done" as one moves up in energy.

It is instructive to understand 
why FCNC $s$ and $b$ decays are so interesting.
Being loop induced, FCNCs are rare because of 
i) a loop factor $\sim g^2/16\pi^2 \sim 10^{-2}$, plus
ii) loop mass (GIM) suppression, which comes in power ($m_i^2/M_W^2$)
and logarithmic ($\log m_i^2/M_W^2$) forms.
A further intrigue from Nature seals the fate for $s$ and $b$ quarks:
the observed mass-mixing hierarchy pattern of
\begin{equation}
 \begin{array}{c}
  m_{u} < m_d \ll m_{s} \ll m_{c} < m_{b} \ll m_{t},  \\
  V_{ub}^2 \ll V_{cb}^2 \ll V_{us}^2 \ll  1.
 \end{array}
\end{equation}
Thus, $c\to s$ and $t\to b$ decays are not suppressed,
while FCNC $c,\ t\to u$, $c$ are 
KM {\it and} loop mass suppressed,
and at best sensitive to genuine BSM effects.
The converse is true for $s$ and $b$:
lifetimes are prolonged by smallness of
$V_{us}$ and $V_{cb}$, $V_{ub}$,
while loop and tree have {\it comparable} KM factors.
In particular, the top drives FCNC $b\to s,\ d$ and 
(CP violating) $s\to d$ processes ({\it penguins!}).
Extending to the hypothetical $b^\prime$,
FCNC decays could dominate its rate.

The one electroweak loop calculations contain vertex
and self-energy diagrams familiar from $g-2$ of QED,
except for the flavor change $Q\to q$,
and one must deal with the complication of several 
(quarks, $W$ and $Z$) masses that must be kept.
For the $K$ and $B$ systems,
the fact that $m_Q^2$, $m_q^2 \ll M_W^2$
allows one to expand in external masses, but keeping 
the internal mass $m_i$ (e.g. $m_t$) dependence exact.\cite{IL}
For the intriguing case of $b^\prime$ decays, 
all loop masses must be kept and 
the calculation is more sophisticated.

For sake of space, we shall not touch 
genuine high scale physics (amply discussed elsewhere in this proceedings), 
CP violation (which often shows up as flavor asymmetries),
and exclusive modes (to avoid hadronic uncertainties).

\section{Rare ({\rm $10^{-10}$ or less}): $\mu$ and the Miraculous $K$}

The abundance of data makes rare $\mu$ decays interesting.
As $\tau_\mu$ is not prolonged, all FCNC effects
such as $\mu\to e\gamma$, $\mu N \to eN$,
$M(\mu^+e^-) \to \bar M(\mu^- e^+)$ are BSM,
with impressive experimental limits that continue to improve.

As for kaons, they  are not only abundant, their lifetimes are 
prolonged by factor of $\vert V_{us}\vert^{-2} \sim 20$.
It is truly remarkable that
the extremely tiny $K_L$--$K_S$ mass difference can be accounted for
by the SM box diagram which is dominated by the $c$ quark.
The genuine FCNC which is CP-conserving is the $K^+\to \pi^+\nu\bar\nu$ mode.
A back-of-envelope estimate is instructive,
\bea
{\rm BR}(K^+\to \pi^+\nu\bar\nu) &\sim& 
{\sum_\nu \vert s\to d\nu\bar\nu\vert^2
 \over \vert s\to ue\bar\nu\vert^2}\, {\rm BR}(K^+\to \pi^0e^+\nu)\nonumber\\
&\sim& 3\times\left\vert \frac{V_{td}V_{ts}}{V_{us}}
\frac{g^2}{16\pi^2}\frac{m_t^2}{M_W^2}
\right\vert^2 \times 0.05 \sim 10^{-10},
\eea
which is still an order of magnitude below the current
limit of $2\times 10^{-9}$, hence a tough experiment indeed.
However, the thunder of the ``first 
penguin" has been stolen by CLEO's observation of $b\to s\gamma$.

\section{Medium ({\rm $10^{-5}$--$10^{-2}$}): the Wonderful $b$}

The $b\to s \gamma$ ($B\to K^*\gamma$) decay can be viewed as
the first ever observed penguin.
The $B$ lifetime is prolonged by $\vert V_{cb}\vert^{-2}\sim 600\times$.
It has become relatively abundant in recent years,
accumulating at the rate of $10^6$ recorded B's per year at present,
hopefully growing to  $10^8/$yr by year 2000 with turn on of B Factories.
Note that the first harbinger for
heavy top and large BCNC effect came from 
the 1987 ARGUS observation of $B$-$\bar B$ mixing
with $\Delta m_{B_d} \sim \Gamma_{B_d}$,
which can be accounted for in SM 
by the box diagram via $m_t$ dominance.

\vskip0.2cm

\noindent (a) \underline{$b\to s\ell^+\ell^-,\ s\nu\bar\nu$}: $Z$ diagram dominance
 \hskip1.3cm $\Longrightarrow$  \hskip1.3cm  $ \sim 10^{-5}$--$10^{-4}$

\vskip0.1cm

Naivly one would expect
$b\to s\gamma^* \to s\ell^+\ell^-$ to dominate over 
$b\to sZ^* \to s\ell^+\ell^-$, since the former is $\sim \alpha G_F$
while the latter is $\sim G_F^2 m^2$.
However, in spontaneously broken gauge theories,
one has non-decoupling of heavy quarks, and the $m^2$ above
turns out to be $m_t^2$ hence $G_F^2 m_t^2 > \alpha G_F$.
A full calculation\cite{HWS}  including box diagrams (which is nothing
but repeating the Inami-Lim results\cite{IL} for $K$ system) confirms this.
The upshot is that the inclusive BR could approach $10^{-5}$,
up from order $10^{-6}$ from photonic penguin alone.
Subsequent detailed work has become an industry.
The physics is rich, and is accessible to experimental
study once one has sufficient rate.

The $b\to s\nu\bar\nu$ process is analogous to $s\to d\nu\bar\nu$, 
with inclusive BR $\sim 10^{-4}$ it is much larger than $10^{-10}$.
This mode, however, is experimentally difficult.

The $b\to s\ell^+\ell^-,\ s\nu\bar\nu$ modes are not sensitive to 
$H^+$ effects, since the dominant $bsZ$ coupling is 
constrained by $\Delta m_{B_d}$.
However, $t^\prime$ effects could be significant if
$V_{t^\prime s} V_{t^\prime b}$ is appreciable.

\vskip0.2cm

\noindent (b)  \underline{$b\to s\gamma$}: Subtle Theory
 \hskip3.05cm $\Longrightarrow$  \hskip3.05cm  $ \sim 10^{-4}$

\vskip0.1cm

From current conservation, one has the effective $bs\gamma$ couplings
\be
F_1 (q^2\gamma_\mu - q_\mu\not{\! q})L + F_2(i\sigma_{\mu\nu}q^\nu m_b)R.
\ee
The effective ``charge radius" term vanishes as $q^2\rightarrow 0$, 
i.e. only the spin-flip ``dipole" transition contributes to
on-shell photonic decay.
However, $F_1$ contains the large-log term of the form
$\log m_i^2/M_W^2$ since it is sensitive to
$b\to s(\bar u u,\ \bar c c) \to sg^*$ on-shell rescattering
when $q^2$ (of $g^*$) is above threshold.
 $F_2$, however, demands an extra spin-flip, 
and suffers from power GIM suppression,
i.e. $\propto m_i^2/M_W^2$.
This leads to an extremely suppressed $b\to s\gamma$.

It was discovered in 1987, however, 
that taking QCD corrections into account
the $b\to s\gamma$ rate is greatly enhanced.\cite{bsA}
Serious calculations using OPE formalism has since become an
industry, now reaching 3 loop order.
The essence, however, can be understood as follows:
$F_2 \sim \alpha_s^0\, (m_i^2/M_W^2 + \cdots) + 
\alpha_s^1\, (\log m_i^2/M_W^2 + \cdots) + \cdots$.
Thus, large-logs appear at $\alpha_s^1$ order,
and because of severe $\alpha_s^0$ order GIM suppression,
the higher order effect dominates!

For the same reason, the  $b\to s\gamma$ $F_2$ coupling 
is sensitive to new physics.\\
\noindent \ $\bullet$ $t^\prime$:  
Besides the SM term $v_c\, \Delta F_2^{ct} \equiv 
V_{cs} V_{cb} (F_2^c -F_2^t)$, 
one has the correction $v_{t^\prime} \Delta F_2^{t^\prime t}$.
Since $t^\prime,\ t$ are both heavy,
$\Delta F_2^{t^\prime t}$ is small.
Good agreement between experiment and SM theory
then implies that $V_{t^\prime s} V_{t^\prime b}$
cannot be large.\\
\noindent \ $\bullet$ $H^+$:  Sensitivity arises again
because of spin-flip subtlety.
In 2HDM with NFC, one has the coupling
$\propto 
\bar u_i[\xi m_i V_{ij} L - \xi^\prime V_{ij} m_j R] d_j\, H^+$,
where $\xi \equiv \cot\beta \equiv v_2/v_1$,
and $\xi^\prime = \xi,\ -1/\xi$ in Model I, II (automatic in MSSM).
As first pointed out by Hou and Willey,\cite{HW,GrWi}
because one needs one power of $m_b$ to account for spin-flip
in $b\to s\gamma$, the $H^+$ correction is
$\Delta F_2^{H^+} \propto \xi^2 f(h) + \xi\xi^\prime\, g(h)$,
where $h= m_t^2/m_{H^+}^2$.
The first term arises from $\not{\! p}_b = m_b$,
while the second term comes from the $tbH^+$ coupling,
hence is $\xi^\prime$-dependent.
Interestingly, then, in the popular Model II where
$\xi\xi^\prime = -1$, the $H^+$ effect is $\tan\beta$-{\it independent,
i.e. always present and always constructive,}\cite{HW}
and would always enhance $b\to s\gamma$ for low $m_{H^+}$.
As emphasized by Hewett\cite{Hewett}
prior to experimental measurement,
the observation of ${\cal B}(b\to s\gamma) = (1- 4) \times 10^{-4}$
by CLEO in 1994 lead to the bound
$m_{H^+} > 260$ GeV, illustrating the power of FCNC search on new physics.

\vskip0.2cm

\noindent c)  \underline{$b\to sq\bar q;\ sg$}: Timelike Gluon Dominance
 \hskip1.85cm $\Longrightarrow$  \hskip1.85cm  $ \sim 1\%$

\vskip0.1cm

It took some time to realize that
the large $m_b$ scale allows the distinction\cite{HSS} of
gluon $q^2$:
lightlike ($b\to sg$, analogous to $b\to s\gamma$),
spacelike ($b\bar q\to s\bar q$, the na\"\i ve extension from $K$-system)
and timelike $b\to sq\bar q$ penguins.
Again because of the smallness of $F_2$, it turns out that 
the timelike penguin dominates and is at the 1\% level.\cite{HSS}
The result is robust against QCD corrections, since the $F_1$ term 
already contains the large-log. 
Exclusive modes such as $B\to K\pi$ are expected at the
$10^{-5}$ level, which are just starting to emerge from CLEO.

The $b\to sq\bar q$ mode is insensitive to $t^\prime$ because of
weak $m_t$, $m_{t^\prime}$ dependence.
However, the $b\to sg$ mode is rather sensitive\cite{HW} to $H^+$,
much like $b\to s\gamma$.
Unfortunately, the rate is highly constrained by $b\to s\gamma$,
and cannot be much larger than the SM result of $\sim 0.2\%$.
Experimentally, however, it could still easily be at $10\%$ order
by some BSM physics and would still go undetected,
but could explain the low semileptonic BR and charm counting rate.

\section{(Pseudo-)Well-Done: FCNC $b^\prime$ Decays May Be Dominant!}

From the rare $10^{-10}$ level in $K$ decay to the
medium $10^{-5}$--1\% BR for $B$ decays, it seems that
{\it FCNC is progressive as one moves up in mass scale.}
This has much to do with the fact that the $d$-type quarks
are lighter than their $u$-type partners,
and their lifetimes are prolonged by CKM suppression in rate.
One naturally turns toward the hypothetical $b^\prime$ system,
where simple extrapolation leads one to expect FCNC
 dominance, which indeed could be the case.

The $b^\prime \to b\gamma,\ bg$ modes were not seen
in a search done by D$\emptyset$, hence $m_{b^\prime} < M_Z$
is ruled out.\cite{HS}
We therefore concentrate on the scenario of
$b^\prime\to bZ,\ bH$ dominance.\cite{HS,HES}
The mechanism is as follows.\cite{HS}
With $m_t \simeq 175$ GeV, there is much room for $m_{b^\prime} < m_t$
hence $b^\prime \not\to t$.
Since $\Gamma(b^\prime \to cW) \propto \vert V_{cb^\prime}\vert^2$,
it could be extremely suppressed if 
$\vert V_{cb^\prime}\vert^2 < \vert V_{ub}\vert^2 \simeq 10^{-5}$.
In comparison, $b^\prime\to bZ,\ bH$ decays
are induced at the loop level precisely by
$t$ and $t^\prime$ intermediate states carrying large
CKM factors $V_{tb^\prime} V_{tb} \simeq - 
V_{t^\prime b^\prime} V_{t^\prime b}$,
while the amplitude is $\propto$ $m_{t^\prime}^2,\ m_t^2$.
Thus, so long that $\vert V_{cb^\prime}/V_{tb^\prime}V_{tb}\vert$
is of order $10^{-2}$ or less, 
$b^\prime\to bZ$ and $b^\prime\to bH$ (when kinematically allowed) decays
dominate over $b^\prime \to cW$.

Such a scenario should not be taken lightly,
even with $N_\nu = 3$ from LEP.
Afterall, the latter only implies $m_{L^0} > M_Z/2$
in the case that the 4th generation does exist.
Searching for $b^\prime$ in the range
$M_Z < m_{b^\prime} < m_t$ via 
$q\bar q,\ gg \to b^\prime \bar b^\prime \to WZ+X$,
$WH + X$, $ZH + X$ and $ZZ + X$ is not only doable at
present Tevatron energies and luminosities,
it might even catch\cite{HS2} the light Higgs boson!
The strong production cross section is certainly orders of
magnitude larger than the standard light Higgs search channel
of $q \bar q^\prime \to WH$ production,
of interest for the future high luminosity option at the Tevatron.

\section{Well-Done: Large FCNC at Weak Scale?}

Although $b^\prime \to b$ could be dominant,
$\Gamma(b^\prime \to b+ X) < 0.1$ MeV is still loop suppressed.
{\it Is it possible for tree level FCNC dominance at high energy?}

Recall the age old problem of family repetition: 
``Who ordered that?".\cite{Rabi}
Since the heady discovery days of 1970-1987, 
we now have the more vexing problem
of mass-mixing hierarchy patterns, Eq. (1).
None of these were anticipated, and together they constitute
the flavor problem.
We simply do not understand the origin of family repetition
and mass-mixing hierarchy pattern.

Besides the flavor problem, the other remaining frontier 
in particle physics is electroweak symmetry breaking.
Although we tend to think that ``we" are ``normal",
yet ``we", you and I, are made of $u$, $d$, $e$ 
(together with solar $\nu_e$) plus the forces. 
Hence, our scale is $ \ll v$.
We might at first think that the top is abnormally heavy since
$m_t \simeq 175$ GeV $\gg m_f,\ \forall f\not = t$.
But, switching Gestalt, we ask:
{\it Is Top \rm (in fact the Only) \it Normal?} That is,
$\lambda_t = \sqrt{2} m_t/v \simeq 1$
is close to the gauge couplings
$g_3 > g_2 > g_1 \sim 0.4$ (at $M_Z$ scale)!

If top  is in fact ``normal", while ``we" are
made of various kinds of ``zero modes",
we would expect {\it new spectra around v},
appearing both in the form of fermions and bosons,
with top still unique since $m_t \gg m_b$.
This motivates us for making further extensions
that break the stranglehold of GIM and NFC.

\vskip0.19cm

\noindent (a)  GIM Breaking: Nonsequential Fermions

\vskip0.02cm

For example, adding singlet quarks $Q_L$ and $Q_R$
breaks GIM and leads to 
tree level FCNC $QqZ^0$ and $QqH^0$ couplings,
which could be much larger\cite{Raj} than loop induced
$b^\prime bZ$ and $b^\prime bH$ couplings.

\vskip0.19cm

\noindent (b)  Foresaking NFC: 2HDM-III

\vskip0.05cm

With two Higgs doublets and without imposing NFC condition,
in general it is impossible to simultaneously diagonalize
the $u$ or $d$ quark mass matrix and their associated two 
Yukawa coupling matrices.
One would thus have tree level FCNH couplings
$f_{ij}$ for each neutral scalar $S^0 = H^0$, $h^0$ and $A^0$.
However, inspired by the mass-mixing hierarchy pattern of Eq. (1),
Cheng and Sher pointed out that FCNH couplings invloving light
quarks are naturally suppressed,\cite{CS} 
e.g. $f_{ij} \sim \sqrt{m_i m_j}/v$,
and could easily evade low energy detection.
An immediate consequence is that
the largest FCNH coupling likely involves the top quark,\cite{Hou,HaWe}
namely, $f_{ct}$, which certainly could be 
larger than $\sqrt{m_c m_t}/v$.
There is practically no experimental limit on
$tcS^0$ couplings.
The resulting phenomenology of 
{\it tree level}\cite{Hou,HaWe} $t\to cS^0$ or\cite{Hou} $S^0\to t\bar c$ decays
is a fascinating subject to be studied at
future colliders such as the LHC, NLC or $\mu^+\mu^-$ collider.
For sake of space, we refer the interested reader to 
current work along these lines.\cite{HLMY}

FCNC  may be ``well-done" at weak scale afterall!?


\section*{References}
\begin{thebibliography}{99} 
%
\bibitem{GW} S.L. Glashow and S. Weinberg, \Journal{\PRD}{15}{1958}{1977}.
\bibitem{IL}T. Inami and C.S. Lim, \Journal{\PTP}{65}{297}{1981}.
\bibitem{HWS}W.S. Hou, R.S. Willey, and A. Soni, \Journal{\PRL}{58}{1608}{1987}.
\bibitem{bsA}S. Bertolini, F. Borzumati, and A. Masiero,
\Journal{\PRL}{59}{180}{1987};
 N.G. Deshpande {\it et al.}, \Journal{\PRL}{59}{184}{1987}.
\bibitem{HW}W.S. Hou and R.S. Willey, \Journal{\PLB}{202}{591}{1988}.
\bibitem{GrWi}See also B. Grinstein and M.B. Wise, \Journal{\PLB}{201}{274}{1988}.
\bibitem{Hewett}J.L. Hewett, \Journal{\PRL}{59}{1521}{1993}; 
\bibitem{HSS}W.S. Hou, A. Soni, and H. Steger, \Journal{\PRL}{59}{1521}{1987}; 
W.S. Hou \Journal{\NPB}{308}{561}{1988}.
\bibitem{HS}W.S. Hou and R.G. Stuart, \Journal{\PRL}{59}{1521}{1989}.
\bibitem{HES}B. Haeri, G. Eilam, and A. Soni, \Journal{\PRL}{59}{1521}{1989}.
\bibitem{HS2}W.S. Hou and R.G. Stuart, \Journal{\PRD}{59}{1521}{1991}.
\bibitem{Rabi}Said to be due to I.I. Rabi but the occasion is not known to me.
\bibitem{Raj}S. Rajpoot, this proceedings.
\bibitem{CS}T.P. Cheng and M. Sher, \Journal{\PRD}{59}{1521}{1987}.
\bibitem{Hou}W.S. Hou, \Journal{\PLB}{59}{1521}{1992}.
\bibitem{HaWe}L. Hall and S. Weinberg, \Journal{\PRD}{59}{1521}{1993}.
\bibitem{HLMY}See, e.g. W.S. Hou {\it et al.}, hep-ph/9702260;
S. Bar-Shalom {\it et al.}, hep-ph/9703221; and references therein.
\end{thebibliography}
\end{document}